\documentstyle[prl,aps,epsf]{revtex}
\begin{document}
\draft \title{Impurities and orbital dependent 
superconductivity in Sr$_2$RuO$_4$} 
\author{D.F. Agterberg}
\address{National High Magnetic Field Laboratory, Florida State University, 
Tallahassee, Florida 32306}
\date{\today}
\maketitle
\begin{abstract}
There now exists a wealth of experimental evidence that
Sr$_2$RuO$_4$ is an odd-parity superconductor. Experiments 
further indicate that among the bands stemming from
the Ru $\{xy,xz,yz\}$ orbitals,  
 the portion of the Fermi surface arising from the
$xy$ orbitals exhibits a much larger gap than the portions
of the Fermi surface arising from the $\{xz,yz\}$ orbitals. 
In this paper 
the role of impurities on such an orbital dependent 
superconducting state 
is examined within the Born approximation.
In contrast to expected results for a nodeless $p$-wave
superconductor the unique nature of the superconducting state
in Sr$_2$RuO$_4$ implies that a low concentration of impurities
strongly influences the low temperature behavior.
\end{abstract}

Since the discovery of superconductivity in the layered oxide Sr$_2$RuO$_4$ 
in 1994 \cite{mae94} and the prediction of odd-parity superconductivity 
\cite{ric95}  
it has been quickly established that the symmetry of the superconducting order
parameter is indeed of odd parity. Early NQR \cite{ish97}, tunneling \cite{jin98},  and impurity studies 
\cite{mac98} clearly indicated
that Sr$_2$RuO$_4$ is not a conventional superconductor. More recently 
$\mu$Sr measurements reveal that the superconducting state breaks time
reversal symmetry \cite{luk98} and Knight shift measurements show no 
change in the spin
susceptibility  when passing through the superconducting transition
\cite{ish98}.
These measurements indicate that the superconducting state is  
described by a spin-triplet pair amplitude with an
orbital dependence $\eta_1 k_x + \eta_2 k_y$ where 
$(\eta_1,\eta_2)\propto(1,i)$ when no magnetic field is applied.  
Such a superconducting state is nodeless in a quasi-2D material.
The effect of impurities within the Born approximation on a nodeless
$p$-wave state has been well studied \cite{gor85,ued85}. 
The results indicate that  
impurities do not drastically change the low temperature properties
of the superconducting state
unless a sufficiently large impurity concentration is present. 
For Sr$_2$RuO$_4$ the experiments of Nishizaki {\it et al.} indicate
that impurities strongly alter the low temperature properties
even in the small impurity concentration limit \cite{nis98}. 
This has previously
been interpreted as an indication that Sr$_2$RuO$_4$ is in the
unitarity scattering limit \cite{mak99}. Here it is shown that the Born
approximation can explain the experimental results once the the unique
microscopic (orbital dependent) nature of the 
superconducting state in Sr$_2$RuO$_4$ is 
considered. 

The superconducting state described above is fully gapped so it
is difficult to understand the experimental observation that  
only approximately half of the Fermi surface exhibited an energy gap
\cite{nis98}.  It was suggested that this feature can be 
understood when the highly planar character and the electronic structure
of the Ru ions are considered  
\cite{agt97}. The formal valence is Ru$^{4+}$
which implies that the electronic properties are due to four electrons 
in bands described by Wannier functions with Ru $d_{xy},d_{xz}$ and $d_{yz}$ 
orbital character. The quasi-2D nature of the electronic dispersion  
and the different parity under the reflection symmetry $\sigma_z$
($z\rightarrow -z$) of the $xy$ and the $\{xz,yz\}$ Wannier functions  
implies that the bands are derived from either the $xy$ or the $\{xz,yz\}$ 
Wannier functions (strictly speaking this is correct only in the 2D limit). 
The parity difference under $\sigma_z$ further inhibits 
Cooper pair scattering between the $xy$ and the $\{xz,yz\}$ bands \cite{agt97}. Consequently, 
the superconductivity in the $xy$ and the $\{xz,yz\}$ bands can be considered
as nearly independent and to a first approximation 
the specific heat data can be understood as the appearance
of superconductivity in either the $xy$ sheet or the $\{xz,yz\}$ sheets     
of the Fermi surface. This theory has experimental support beyond 
the specific heat measurements.  In particular     
which band is responsible for the superconductivity 
has been addressed experimentally by Riseman {\it et al.} using 
small angle neutron scattering \cite{for98}. For the superconducting state with 
the two component order parameter described above 
it has been predicted that a square vortex lattice 
is expected to occur when the field is applied along the $c$-axis \cite{agt98}.
The orientation of the square vortex lattice depends upon which of the 
bands are involved in the superconductivity.
Riseman {\it et al.} observed a square vortex lattice and its orientation
implies that the $xy$ sheet of the Fermi surface is superconducting 
\cite{for98}. Furthermore, the measured size of the penetration depth
is consistent with pairing on the $xy$ sheet of the Fermi surface 
but not consistent with pairing over the whole Fermi surface \cite{for98}.
It is of interest to note as well that the measurements of Imai {\it et al.}
indicate that only the spin susceptibility of the $xy$ orbitals   
exhibits a significant increase with decreasing temperature \cite{ima98}.
 
In this paper the role
of impurities on orbital dependent superconductivity will be 
considered within the Born approximation. 
A standard approach using a nodeless $p$-wave state
does not explain  the rapid increase in the residual 
density of states with impurity 
concentration 
as seen by Nishizaki \cite{nis98}. 
For this reason Maki and Puchkaryov proposed 
that the impurities in Sr$_2$RuO$_4$ are
sufficiently strong scatterers that the unitarity limit of the impurity 
problem should be used \cite{mak99}.
The unitarity limit has also been argued to be relevant for
heavy fermion systems \cite{sch86,hir86} and for high $T_c$ materials
\cite{hir93,hot93,sun95,hot95}. 
In the heavy fermion
case this limit is plausible due to small Fermi temperature ($\approx 10$ K) 
that occurs in these materials while for high-$T_c$ materials the filled
$d$-shell of Zn impurities  
may justify the unitarity limit. However neither
of these two plausibility arguments apply to Sr$_2$RuO$_4$ so, while the 
unitarity limit cannot be ruled out,  it is worthwhile considering the 
Born limit in more detail. 
Here it is shown that a more realistic treatment of the  
impurity scattering within the Born approximation 
and within the context of orbital dependent superconductivity can explain
the experimental results. The key element is inter-band scattering
in which a quasiparticle with even parity under $\sigma_z$ is scattered 
to a quasiparticle state with odd-parity under $\sigma_z$. Such
a scattering implies that reflection under $\sigma_z$ is locally broken    
by the impurity. This can occur for example if the impurity does not lie in
the RuO$_4$ plane, if it induces
a rotation of the RuO$_6$ octahedra about an in plane axis, or if there are
layer stacking defects.  
After averaging over all impurity positions, 
this inter-band scattering does not mix the 
quasiparticle states of different
parity, but does affect decrease the quasiparticle lifetimes.
Here calculations of the specific heat as a function of temperature
are performed to examine the consequences of this model and to 
compare to existing experimental data.

LDA band structure calculations of Sr$_2$RuO$_4$
\cite{ogu95,sin95} reveal that the
density of states
near the Fermi surface is due mainly to the four Ru $4d$ electrons
in the $t_{2g}$ orbitals.
There is a strong hybridization of these orbitals with the O
$2p$ orbitals giving rise to antibonding $\pi^*$ bands. The resulting
bands have three quasi-2D Fermi surface sheets labeled $\alpha,
\beta,$
and $\gamma$ (see Ref. \cite{mac296}). In the planar limit due to the 
different parity under reflection ($z\rightarrow -z$) the 
$\alpha$ and $\beta$ sheets
consist solely of $\{xz,yz\}$ Wannier functions and the
$\gamma$ sheet of $xy$ Wannier functions. 
I assume an impurity potential
that obeys the symmetry relation $V({\bf r})=V(\sigma_z{\bf r})$ where
$\sigma_z$ is reflection through the Ru$_2$O$_4$ plane and $r_z=0$ lies
in the Ru$_2$O$_4$ plane. 
On average such an impurity potential will not mix 
single particle excitations corresponding to 
different parity under $\sigma_z$.
 However the single particle excitations on each sheet will have 
two contributions to the lifetime: one from eigenstates of the same parity
(intra-band)
and one from eigenstates of opposite parity (inter-band) under $\sigma_z$.
In the model considered below it will be assumed that the $\alpha$ and
$\beta$ sheets are equivalent with respect to the single particle and
superconducting properties. This is correct if a nearest neighbor
tight binding dispersion is used to describe these sheets and is a
reasonable  approximation for more realistic dispersion relations. 
This leads to an effective two band model for the superconducting state
in which the eigenstates of each band have opposite parity under $\sigma_z$.  
For simplicity two cylindrical Fermi surface sheets will be used: one
with density of states equal to that of the $\gamma$ sheet and one with density
of states equal to that of the $\alpha$ and $\beta$ sheets (based on the
measurements of Ref.~\cite{mac296} I take 
$N_{\gamma}:(N_{\alpha}+N_{\beta})$ to be $0.55:0.45$). The 
interaction leading to superconductivity is taken to have the form
\begin{equation}
V_{l,l^{\prime}}({\bf k},{\bf k}^{\prime})=V_{l,l^{\prime}}
\frac{
{\bf k}\cdot{\bf k}^{\prime}}{k_{F_l}k_{F_{l^{\prime}}}}
\end{equation}
where $k_{F_l}$ is the magnitude of the Fermi wavevector on sheet $l$. 
The gap matrix
on each sheet is then of the form 
\begin{equation}
\hat{\Delta}(l,{\bf k})=\pmatrix{0&c_l(k_x+ik_y)/k_{F_l}\cr
c_l(k_x+ik_y)/k_{F_l}&0}.
\end{equation}
Only the $s$-wave scattering potential is included and the $s$-wave scattering
between band $l$ and $l^{\prime}$ is characterized by
$u_{l,l^{\prime}}$. The resulting gap
and self-energy equations are
\begin{equation}
c_l=T\pi\sum_{l^{\prime},n}\frac{\tilde{V}_{l,l^{\prime}}c_{l^{\prime}}}
{\sqrt{\tilde{w}_{n,l^{\prime}}^2+c_{l^{\prime}}^2}}
\end{equation}
and 
\begin{equation}
\tilde{w}_{n,l}=w_n+\sum_{l^{\prime}}\frac{\Gamma_{l,l^{\prime}}
\tilde{w}_{n,l^{\prime}}}{\sqrt{\tilde{w}_{n,l^{\prime}}^2+c_{l^{\prime}}^2}}
\end{equation}
where $\tilde{V}_{l,l^{\prime}}=N_{l^{\prime}}V_{l,l^{\prime}}$,
$\Gamma_{l,l{^\prime}}=\pi n_i N_{l^{\prime}} [u_{l,l^{\prime}}]^2$,
$N_l$ is the normal density of states on sheet $l$, 
and $n_i$ is the concentration of impurities.
Note that due to the odd-parity symmetry of the order parameter there is no
re-normalization of the gap,
consequently strong inter-band scattering does not lead to a state
with an equal gap over the whole Fermi surface 
as is the case in $s$-wave
superconductors \cite{gol97,sch77}. This is partially a consequence  
of keeping only the $s$-wave Born scattering amplitude. If anisotropic
contributions to the inter-band scattering amplitude are also included 
then in principle these terms will induce a gap on the $\{\alpha,\beta\}$
sheets of the Fermi surface (if none exists in the clean limit).  
However this induced gap is subject to pair breaking due to the 
$s$-wave scattering amplitudes.  This indicates that provided the anisotropic 
scattering amplitude is not too large keeping only the $s$-wave scattering
amplitudes will capture the
underlying physics of Sr$_2$RuO$_4$ 
(see References \cite{haa97,har98} for a discussion
of physical consequences of anisotropic impurity potentials). 
 
The equation for $T_c$ deviates slightly from the standard
Abrikosov-Gor'kov form and is given by
\begin{eqnarray}
\ln(T_c/T_c^0)=&f_+-\frac{\sqrt{(V_{11}-V_{22})^2+4V_{12}V_{21}}}{2(V_{11}
V_{22}-V_{12}V_{21})}\\
&+\frac{\sqrt{(V_{11}+V_{22})^2-4[1-f_-^2(V_{11}V_{22}
+V_{12}V_{21})-f_-(V_{11}-V_{22})](V_{11}V_{22}-V_{12}V_{21})}} 
{2(V_{11}
V_{22}-V_{12}V_{21})}
\end{eqnarray}
where $f_+=[\Psi(\frac{1}{2}+\rho_1)+
\Psi(\frac{1}{2}+\rho_2)]/2 -\Psi(\frac{1}{2})$, $f_-=[\Psi(\frac{1}{2}+\rho_1)
-\Psi(\frac{1}{2}+\rho_2)]/2$,  $\Psi(x)$ is the digamma function,
$\rho_i=(\Gamma_{i1}+\Gamma_{i2})/(2\pi T_c)$, and $T_c^0$ is the transition
temperature in the presence of no impurities.

\begin{figure}
\epsfxsize=80mm
\centerline{\epsffile{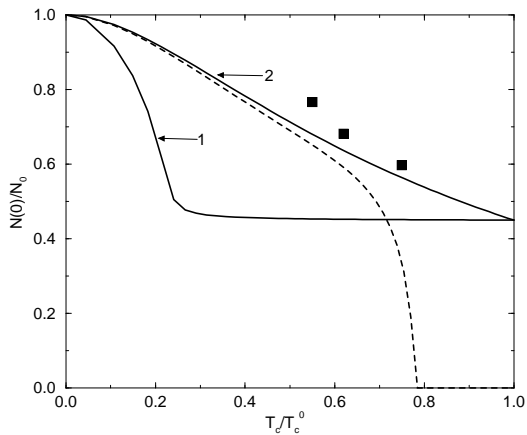}}
\caption{Residual density of states for different impurity
concentrations. The curve labeled 1 (2) is in the strong 
intra-band (inter-band) scattering limit and the experimental 
estimates are shown as squares. The dotted curve is in the
strong inter-band scattering limit when the gap on the 
$\{\alpha, \beta\}$ sheets is one tenth that of the $\gamma$ sheet.}     
\label{fig1}
\end{figure}

The density of states (DOS) is given by 
\begin{eqnarray}
N(w)=&-\frac{1}{\pi}\sum_l\int\frac{d^2k}{(2\pi)^2}\Im [G_l(k,z)|_{z\rightarrow w+
i\delta}]\\
&=\sum_lN_l\sum_{l^{\prime}}[\Gamma^{-1}]_{l,l^{\prime}}\Im (\tilde{z}_{l^{\prime}})
\end{eqnarray} 
It is of interest to determine $N(w=0)$ in the limit of zero temperature as
this quantity is measurable as the residual DOS in specific
heat measurements. To date there has been no experimental evidence for a gap
appearing on the $\alpha$ and $\beta$ sheets of the Fermi surface in 
Sr$_2$RuO$_4$. In view of this I consider initially the limit $c_2=0$.
In Fig.~\ref{fig1} the residual DOS is plotted as a function of the transition 
temperature for the strong inter-band and intra-band scattering limits.  
Also shown in this Figure is the extrapolation from $T=0.3$ K of the same 
quantity from the data of Nishizaki {\it et al.} \cite{nis98} 
(in Fig.~\ref{fig1} $T_c^0$ has been assumed to be 1.5 K).   
While a detailed comparison to the experimental data will require 
measurements at lower temperature it is clear 
that the intra-band scattering limit cannot account for 
the data. This  indicates that inter-band scattering cannot be neglected.
Note that in the limit $\Gamma_{ii}=0$ and $c_2=0$ the density of states
can be found analytically:
\begin{equation}
N_1(w)=
\frac{w\sqrt{\sqrt{(c_1^2+\Gamma_{12}^2-w^2)^2+4
\Gamma_{12}^2w^2}+w^2-c_1^2-\Gamma_{12}^2} +\Gamma_{12}
\sqrt{\sqrt{(c_1^2+\Gamma_{12}^2-w^2)^2+
4\Gamma_{12}^2w^2}+c_1^2+\Gamma_{12}^2-w^2}}
{\sqrt{2}\sqrt{(c_1^2+\Gamma_{12}^2-w^2)^2+
4\Gamma_{12}^2w^2}}
\end{equation}
In the zero frequency limit 
$N(0)=N_2+\frac{\Gamma_{12}}{\sqrt{c_1^2+\Gamma_{12}^2}}$ 
showing that inter-band scattering  increases 
$N(0)$ from $N_2$ for infinitesimal $\Gamma_{12}$ which gives rise to the
residual DOS seen in Fig.~\ref{fig1} 
Also shown in Fig.~\ref{fig1} is the residual density of states when $c_2=c_1/10$
in the strong inter-band scattering limit.

\begin{figure}
\epsfxsize=80mm
\centerline{\epsffile{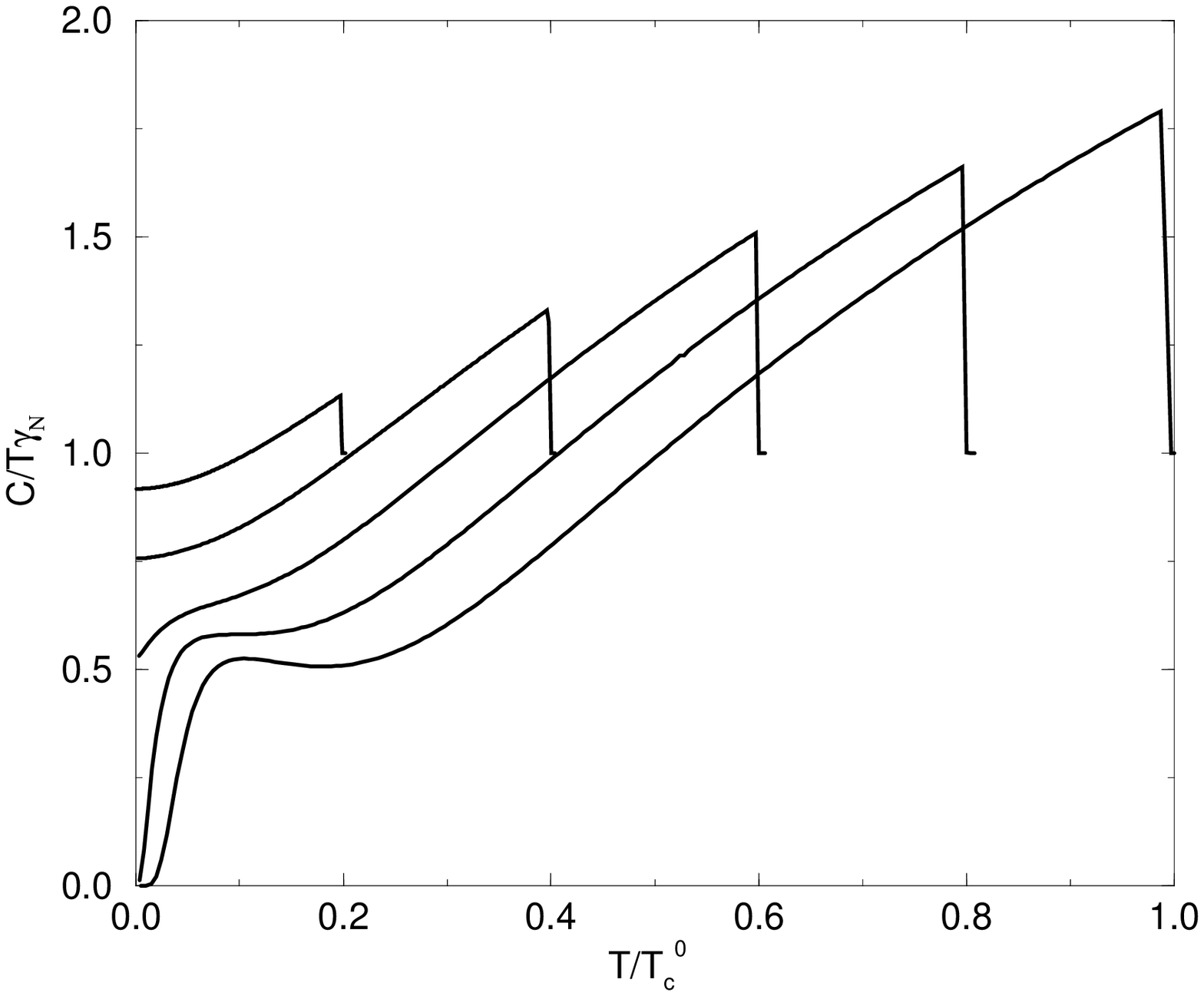}}
\epsfxsize=80mm
\centerline{\epsffile{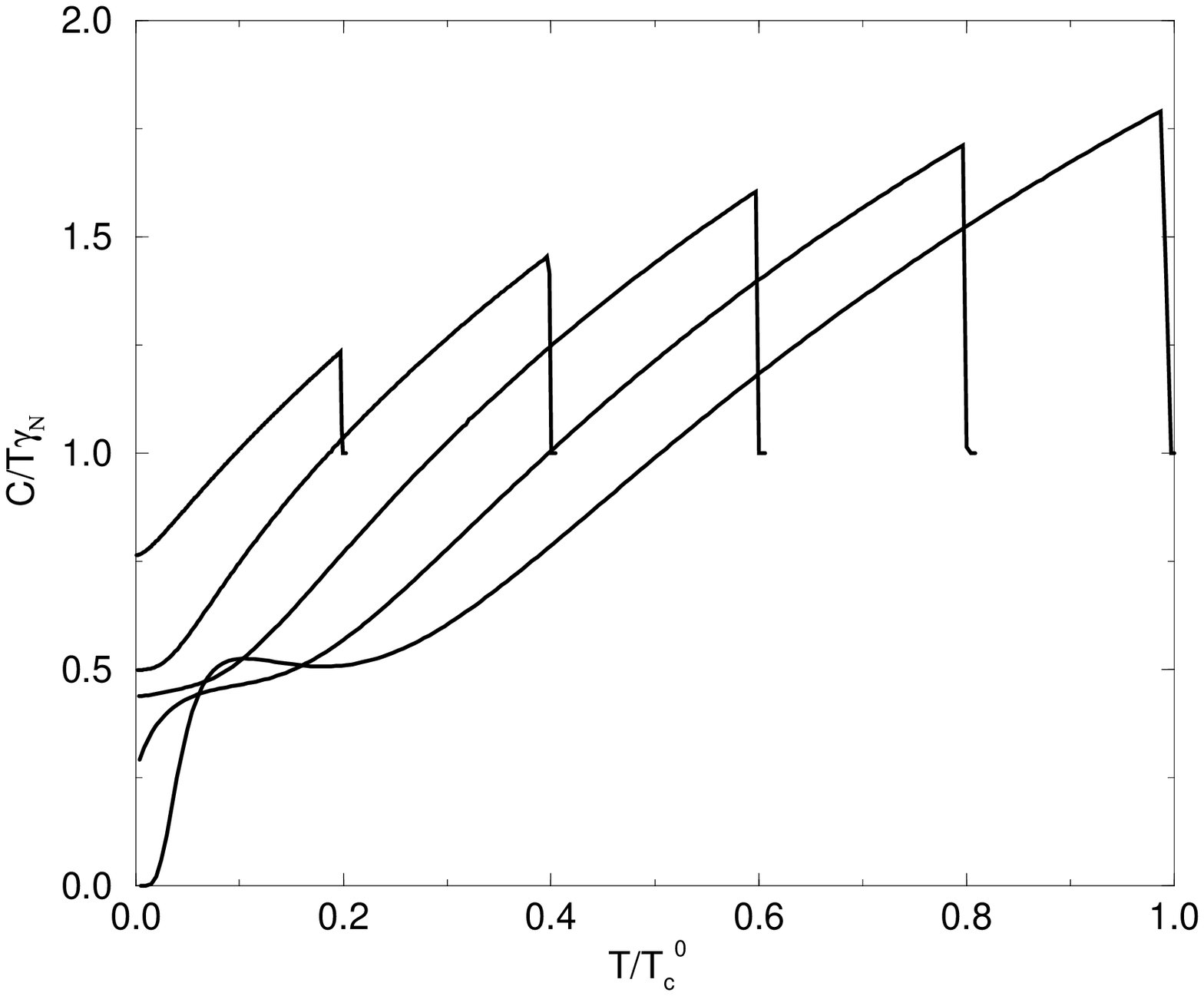}}
\caption{Evolution of the $C/T$ with increasing impurity concentration.  
The top (bottom) Figure is in the the strong intra-band (inter-band) 
scattering limit.}
\label{fig2}
\end{figure}

The specific heat is calculated by numerically evaluating the 
temperature derivative of the entropy. The entropy is given by
\begin{equation}
S=\int_0^{\epsilon_c} d\epsilon N(\epsilon)\{[1-f(\epsilon)]\ln[1-f(\epsilon)]+
f(\epsilon)\ln f(\epsilon)\}.
\end{equation}
where $\epsilon_c$ is the cut-off energy for the BCS interaction
and $f(\epsilon)$ is the Fermi distribution function.
In Fig.~\ref{fig2}  the results for $C/T$ are shown 
for the intra-band scattering limit ($u_{11}=u_{22}=10 u_{12}$) 
and the inter-band scattering limit ($u_{11}=u_{22}=u_{12}/10$)
in the limit $V_{11}=10 V_{12}$ and $V_{22}=0$.
Fig.~\ref{fig2} shows that inter-band scattering raises the apparent residual
DOS ({\it i.e.} that found by using entropy balance to extrapolate 
from measurements taken down to $T=T_c^0/5$) and decreases
the specific heat jump relative to intra-band scattering. Also   
any gap on the $\{\alpha, \beta\}$ bands will be less rapidly destroyed
by inter-band scattering than by intra-band scattering.  

A theory of impurities within the context of orbital dependent 
superconductivity has been developed for Sr$_2$RuO$_4$. 
In contrast to known results for nodeless $p$-wave superconductors,
a strong dependence of the low temperature behavior of the
specific heat on impurity concentration is shown to be a consequence
of this theory in the Born approximation.  
This behavior is due to  
the impurity scattering between bands of opposite
parity symmetry under reflection through the RuO$_4$ plane. 
This theory accounts for
the recent measurements of the specific heat for different impurity
concentrations provided this inter-band scattering is not neglected.

I wish to thank G. Cao, S. Haas, T. Hotta, A.P. Mackenzie, 
Y. Maeno, K. Maki, I.I. Mazin, V. Mineev, S. Nishizaki, T.M. Rice, and 
M. Sigrist
for useful discussions. 
I acknowledge support from the State of Florida and NSF grant DMR9527035.

\end{document}